\documentclass[conference]{IEEEtran}
\IEEEoverridecommandlockouts
\usepackage{enumitem}
\usepackage{cite}
\usepackage{amsmath,amssymb,amsfonts}
\usepackage[acronym,toc,shortcuts]{glossaries}
\usepackage{enumitem}
\usepackage{graphicx}
\usepackage{textcomp}
\usepackage{xcolor}
\usepackage{csquotes}
\usepackage{balance}
\usepackage{amsmath}
\usepackage{algorithm}
\usepackage{algpseudocode}
\def\BibTeX{{\rm B\kern-.05em{\sc i\kern-.025em b}\kern-.08em
    T\kern-.1667em\lower.7ex\hbox{E}\kern-.125emX}}

\usepackage[acronym,toc,shortcuts]{glossaries}

\graphicspath{{figures/}}

\makeglossaries
\newacronym{3GPP}{3GPP}{The 3rd Generation Partnership Project }
\newacronym{5G}{5G}{Fifth Generation}
\newacronym{AAA}{AAA}{Authentication, Authorization and Accounting}
\newacronym{ARIMA}{ARIMA}{AutoRegressive Integrated Moving Average}
\newacronym{CPU}{CPU}{Central Processing Unit}
\newacronym{GRU}{GRU}{Gated Recurrent Unit}
\newacronym{O-RAN}{O-RAN}{Open Radio Access Network}
\newacronym{QoS}{QoS}{Quality of Service}
\newacronym{PRB}{PRB}{Physical Resource Block}
\newacronym{rApp}{rApp}{radio App}
\newacronym{RAN}{RAN}{Radio Access Network}
\newacronym{RIC}{RIC}{RAN Intelligent Controller}
\newacronym{xApp}{xApp}{eXtended application}
\newacronym{SFF}{SFF}{Simple-Feed-Forward}
\newacronym{LSTM}{LSTM}{Long-Short Term Memory}
\newacronym{SN}{SN}{Seasonal-Naive}
\newacronym{MLP}{MLP}{ Multi-Layer Perceptron}
\newacronym{RNN}{RNN}{Recurrent Neural Network}
\newacronym{MSE}{MSE}{Mean Square Error}
\newacronym{MASE}{MASE}{Mean Absolute Scaled Error}
\newacronym{MAPE}{MAPE}{Mean Absolute Percentage Error}
\newacronym{ND}{ND}{Normalized Deviation}
\newacronym{QL}{QL}{Quantile Loss}
\newacronym{AI}{AI}{Artificial Intelligence}
\newacronym{ML}{ML}{Machine Learning}
\newacronym{SLAs}{SLAs}{Service Level Agreements}
\newacronym{MANO}{MANO}{Management and Orchestration}
\newacronym{OAM}{OAM}{Operations, Administration, and Maintenance}
\newacronym{CAPEX}{CAPEX}{Capital Expenditure}
\newacronym{OPEX}{OPEX}{Operating Expense}
\newacronym{O-RU}{O-RU}{Open-Radio Unit}
\newacronym{O-DU}{O-DU}{Open-Distribution Unit}
\newacronym{O-CU}{O-CU}{Open-Central Unit}

\usepackage{fancyhdr}
\usepackage{geometry}
\usepackage{lipsum} 

\begin{document}

\title{Minimizing Power Consumption under SINR Constraints for Cell-Free Massive MIMO in O-RAN\\
}


\author{Vaishnavi Kasuluru, Luis Blanco, Miguel \'Angel V\'azquez, Cristian J. Vaca-Rubio, Engin Zeydan\\
{\normalsize{} Centre Technologic de Telecomunicacions de Catalunya (CTTC/CERCA), Castelldefels, Barcelona, Spain, 08860.} \\
{\normalsize{} Emails: \texttt{\{vkasuluru, lblanco, mavazquez, cvaca, ezeydan\}@cttc.es}}

\thanks{This work has been supported by SEMANTIC project, funded by the European Union’s Horizon 2020 research and innovation program under the Marie Skłodowska-Curie grant (agreement No 861165), the Spanish projects FREE6G-RadEdge (TSI-063000-2021-121) and FREE6G-RegEdge (TSI-063000-2021-144) funded by MINECO through the “NextGenerationEU” program, and the Spanish project ORIGIN (PID2020-113832RB-C22) funded MICCIN.}

}

\maketitle

\begin{abstract}

This paper deals with the problem of energy consumption minimization in Open RAN cell-free (CF) massive Multiple-Input Multiple-Output (mMIMO) systems under minimum per-user signal-to-noise-plus-interference ratio (SINR) constraints. Considering that several access points (APs) are deployed with multiple antennas, and they jointly serve multiple users on the same time-frequency resources, we design the precoding vectors that minimize the system power consumption, while preserving a minimum SINR for each user. We use a simple, yet representative, power consumption model, which consists of a fixed term that models the power consumption due to activation of the AP and a variable one that depends on the transmitted power. The mentioned problem boils down to a binary-constrained quadratic optimization problem, which is strongly non-convex. In order to solve this problem, we resort to a novel approach, which is based on the penalized convex-concave procedure. The proposed approach can be implemented in an O-RAN cell-free mMIMO system as an xApp in the near-real time RIC (RAN intelligent Controller). Numerical results show the potential of this approach for dealing with joint precoding optimization and AP selection. 

\end{abstract}


\section{Introduction}
\label{intro}

Cell-free (CF) massive multiple-input-multiple-output (mMIMO) systems is a key enabling technology for 6G networks \cite{9356519} seemingly merging and pushing forward two key techniques already in use in 5G, namely, network densification and massive MIMO antenna arrays. In a nutshell, CF-mMIMO takes advantage of network MIMO by assuming a large number of access points (APs)
deployed over the coverage area but with joint baseband processing done by a central processing unit \cite{7827017}. In fact, CF-mMIMO is often described as a conventional mMIMO network whereby the RF heads are pulled apart from the BS and randomly (and densely) scattered throughout the coverage area. The theoretical foundations underpinning mMIMO mostly apply to the CF-mMIMO scenario. In particular, the so-called channel hardening that virtually eliminates the fast fading effects also applies to the CF-mMIMO scenario \cite{10437450}. Furthermore, its highly distributed topology brings the RF infrastructure geographically closer to the users, thus effectively reducing the propagation losses. This theoretical gain is obtained at the expense of costly and power-hungry equipment that connects APs in-phase and quadrature (IQ) signals to the central control unit and the requirement to maintain strict synchronization among all APs.
 
Aligned with the Green Deal initiative promoted by the European Union, energy efficiency will again be a fundamental metric in the forthcoming wireless generation\cite{9353695}. While power consumption in mMIMO mostly addressed issues on radiofrequency and baseband processing units \cite{10437450}
, CF-mMIMO must consider new energy-related aspects such as the presence of a dedicated fronthaul connection linking the APs to the CPU 
and the fixed power expenditure each individual AP entails. In practical networks, these fixed power terms represent the most significant factor in total energy expenditure
. In this context, to reduce global power consumption, it might be wise to just activate a few APs in order to attend to specific user data rate demands. Techniques to dynamically switch on/off APs in an attempt to maximize energy efficiency have been proposed in \cite{femenias2020access} whereby a zero-forcing (ZF) precoder is combined with the heuristic power allocation. 
Unfortunately, the maximization of energy efficiency is carried out without providing the users with any performance guarantees. Most previous works on CF-mMIMO separate the precoder design from the power allocation strategy. While the first one is carried out relying on instantaneous channel state information (CSI), the second one typically targets a specific performance objective such as maximizing the minimum user signal-to-noise-plus-interference ratio (SINR) (Max-Min) or maximizing the network sum rate. 

It has been recently proposed the use of CF-mMIMO within the Open Radio Access Network (O-RAN) architecture \cite{Demir_CF_ORAN}, representing a groundbreaking convergence of two disruptive technologies. Open RAN's disaggregated framework fosters network operator inter-operability, driving down costs and unlocking vendor diversity. When coupled with CF-mMIMO, this synergy amplifies the benefits by breaking down traditional cell boundaries and leveraging distributed antennas, maximizing spectral efficiency, extending coverage, and minimizing interference. This potent combination not only optimizes resource utilization but also accelerates the deployment of advanced wireless networks, empowering operators to meet the ever-growing demands for connectivity.

In this paper, we focus on the optimization problem of minimizing the power consumption of CF-mMIMO systems within O-RAN architectures, under minimum quality-of-service (QoS) requirements per user in the form of a minimum SINR. Indeed, relying on the seminal work in \cite{zhou2020max}, we revisit the idea of minimizing the transmit power under minimum SINR \cite{zhou2020max} for CF-mMIMO considering the fixed power consumption term. In our case, and similar to what happens in \cite{zhou2020max}, the precoder design and power allocation steps merge, and they are both carried out taking into account instantaneous CSI, thus allowing the satisfaction of instantaneous SINR constraints (rather than average). When only the transmit power consumption is assumed, the optimization problem has a convex reformulation \cite{zhou2020max} that permits a closed-form solution. However, incorporating the fixed term transforms the problem into a highly non-convex problem. In order to tackle this new optimization problem, binary decision variables are introduced that decide which APs are active, along with the derivation of the corresponding precoding vectors. We propose a convex relaxation based on the Penalty Convex-Concave Procedure (PCCP) \cite{lipp2016} able to handle the described problem. The numerical results show the benefits of this technique for joint precoding design and AP subset selection, which minimize global power consumption while satisfying user demands.

The paper's outline is as follows: Section II introduces the architecture of the Cell-Free O-RAN system, followed by the System model in Section III. Section IV describes the considered problem statement. Section V and Section VI present the numerical results and conclusions, respectively.

\section{Cell-Free O-RAN Architecture}

\begin{figure*}[htp!]
\centering
\includegraphics[width=0.8\linewidth, height=0.5\textheight]{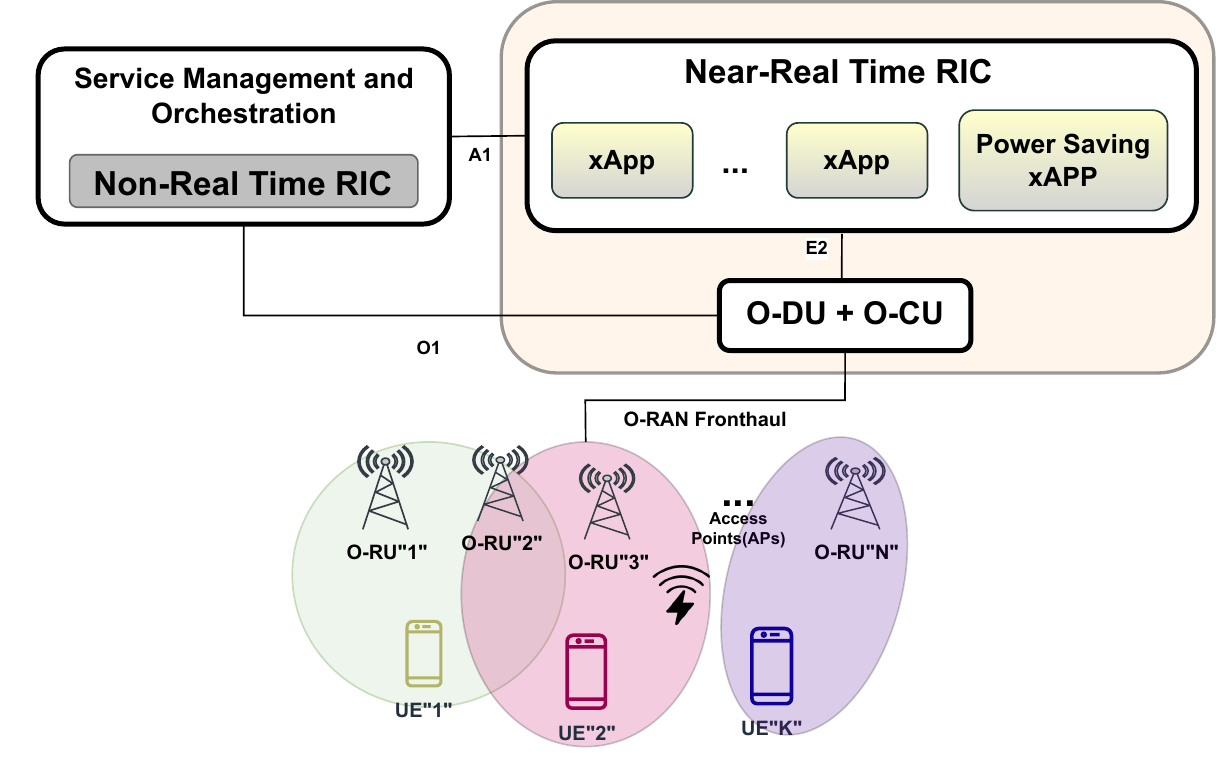}
\caption{Cell-free O-RAN Architecture.}
\label{O-RAN}
\end{figure*}

Traditional RANs, with their hardware dependency, have long posed challenges for network providers. These systems often lead to vendor lock-in issues, significantly increasing CAPEX and OPEX costs \cite{10266607}. The struggle to incorporate intelligence and build a collaborative, reliable network environment has been a major hurdle \cite{polese2023understanding}. The promise of next-generation RAN solutions, which can manage the network in real-time and respond quickly to scenarios using software-defined vendor-neutral technologies, is a hope in this landscape \cite{kliks2023towards}. 

With Virtualization and RAN disaggregation being the cornerstone, O-RAN technology aims to provide openness and intelligence in network management\cite{bonati2021intelligence}. One of the latest essential foundations is the modernization of O-RAN architecture by integrating it with cell-free technology \cite{Demir_CF_ORAN}. This helps in improving the scalability and flexibility of the network. Figure \ref{O-RAN} shows the architecture of O-RAN Cell-free technologies \cite{Demir_CF_ORAN}, incorporating power-saving \ac{xApp}. It consists of control, management, and data layers. In the Management plane, there is a Service Management and Orchestration (SMO) layer equipped with Non-Real Time \ac{RIC}. The Non-RT \ac{RIC} constitutes a fundamental component of the Open RAN architecture, specifically designed to manage RAN operations with control loops exceeding one second. It supports third-party applications, referred to as \ac{rApp}s, which facilitate the optimization and management of RAN functions. As an integral part of the SMO framework within the O-RAN architecture, the Non-RT \ac{RIC} connects to other network elements through A1, O1, and O2 interfaces.

There is also a Near-Real Time \ac{RIC}, which helps in intelligent system management through \ac{xApp}s. In this work, these \ac{xApp}s are considered to use the information from \ac{O-RU}s to make decisions regarding power saving by using a sophisticated optimization algorithm. The \ac{xApp}s primary objective is to promote power efficiency within the network. Moreover, The \ac{O-RU}s are designed to support a cell-free architecture for serving multiple User Equipments (UEs) in a seamless manner. The Near-Real Time \ac{RIC} interfaces via A1 and E2 interfaces with other network elements. 

Combining O-RAN and cell-free technologies provides various benefits. It reduces signal interference and improves the efficient usage of power. As the APs cooperatively serve multiple UEs, the traffic demands can be handled dynamically, which in turn enhances capacity and coverage \cite{Demir_CF_ORAN}.  Nevertheless, it also helps ensure network services' consistent availability to improve user experience and handle scalability and agility cost-effectively in the long run \cite{10437450}. The main components are: 

\begin{itemize}[]
    \item \ac{O-RU}, \ac{O-DU}, and \ac{O-CU} whose functionalities are similar to that in 5G dis-aggregated RAN except with added support of O-RAN based specifications and interface. Without loss of generality, throughout this paper, we will refer to the \ac{O-RU}s as cell-free APs \cite{10330597}.
    \item Near-Real Time \ac{RIC}  to control/optimize RAN elements and resources based on fine-grained data using online \ac{AI}/\ac{ML} based services. It is suitable for applications with latency requirements between 10 ms and 1s \cite{li2017intelligent}. Herein, the power saving \ac{xApp} is placed in Near-Real Time \ac{RIC}, selecting the best subset of APs to be active at a particular time to handle the service requirements of users efficiently.
    \item Non-Real Time \ac{RIC} to control/optimize RAN elements and resources based on coarse-grained data using online \ac{AI}/\ac{ML} services \cite{li2017intelligent}. It is suitable for applications with latency requirements greater than 1s. It also provides policy-based guidance to near-real Time \ac{RIC}.  
\end{itemize}

\section{System Model}

Consider a cell-free scenario with $K$ User terminals / User Equipments (UEs) and $N$ Access Points (APs). In the following, we assume that APs are equipped with $N_{\mathrm{AP}}>1$ antennas. APs are connected through a high-speed link to a central controller. We assume that UEs are equipped with a single antenna.

\subsection{Channel Model}

Let us denote by $\beta_{k}^{[n]}$ the large-scale propagation losses (i.e., path loss and shadowing) of the link joining $\mathrm{AP} n$ and UE $k$, and which can be expressed as $\beta_{k}^{[n]}=\zeta_{k}^{[n]} \chi_{k}^{[n]}$ with $\zeta_{k}^{[n]}$ representing the distance-dependent path loss

\begin{equation*}
\zeta_{k}^{[n]}[\mathrm{dB}]=\zeta_{0}+10 \alpha \log _{10}\left(d_{k}^{[n]}\right), \tag{1}
\end{equation*}
where $\zeta_{0}$ is the path loss at a reference distance of $1 \mathrm{~m}$, $d_{k}^{[n]}$ is the distance from AP $n$ to UE $k$ and $\alpha$ is the path loss exponent. The term $\chi_{k}^{[n]}$ corresponds to the shadow fading component modeled as a spatially correlated log-normal random variable with zero mean and variance $\sigma_{\chi}^{2}$ whose spatial correlation model is described in [\cite{7827017}, (54)-(55)]. The resulting downlink channel vector $\boldsymbol{h}_{k}^{[n]} \in \mathbb{C}^{N_{\mathrm{AP}} \times 1}$ from the $k$-th UE to the $n$-th AP (including both large-scale and small-scale fading) can then be generically characterized as a correlated Rayleigh fading channel $\boldsymbol{h}_{k}^{[n]} \sim \mathcal{C N}\left(\mathbf{0}, \boldsymbol{R}_{n k}\right)$ with $\boldsymbol{R}_{n k}$ denoting the $N_{A P} \times N_{A P}$ spatial channel correlation matrix subject to the constraint $\operatorname{Tr}\left(\boldsymbol{R}_{n k}\right)=\beta_{k}^{[n]} / N^{\mathrm{AP}}$. For later convenience, we define $\boldsymbol{h}_{k}=\left[\boldsymbol{h}_{k}^{[1], T}, \cdots, \boldsymbol{h}_{k}^{[n], T}\right]^{T}$ as the $N_{\mathrm{AP}} \times 1$ vector collecting the overall channel responses from the $N$ APs to user $k$.

\subsection{Uplink training and channel estimation}

During the UL training phase, all $K$ UEs simultaneously transmit pilot sequences of $\tau_{p}$ samples to the APs and thus, the $N_{A P} \times \tau_{p}$ received UL signal matrix at the $n$-th active AP is given by

\begin{equation*}
\boldsymbol{Y}_{p}^{[n]}=\sqrt{\tau_{p} P_{p}^{\mathrm{AP}}} \sum_{k=1}^{K} \boldsymbol{h}_{k}^{[n]} \boldsymbol{\varphi}_{k}^{T}+\boldsymbol{N}_{p}^{[n]} \tag{2},
\end{equation*}

where $P_{p}^{\mathrm{AP}}$ is the available pilot symbol power, $\varphi_{k}$ denotes the $\tau_{p} \times 1$ training sequence assigned to UE $k$, with $\left\|\varphi_{k}\right\|^{2}=1$, and $N_{p}{ }^{[n]} \in \mathbb{C}^{N_{\mathrm{AP}} \times \tau_{p}}$ is a matrix of independent identically distributed (iid) zero-mean circularly symmetric Gaussian random variables with standard deviation $\sigma_{u}$. Define now the projection of $\boldsymbol{Y}_{p}{ }^{[n]}$ on the $k$-th training sequence

\begin{equation*}
\breve{\boldsymbol{y}}_{p k}^{[n]}=\boldsymbol{Y}_{p}^{[n]} \boldsymbol{\varphi}_{k}^{*}=\sum_{k^{\prime}=1}^{K} \sqrt{\tau_{p} P_{p}^{\mathrm{AP}}} \boldsymbol{h}_{k^{\prime}}^{[n]} \boldsymbol{\varphi}_{k^{\prime}}^{T} \boldsymbol{\varphi}_{k}^{*}+\boldsymbol{N}_{p}^{[n]} \boldsymbol{\varphi}_{k}^{*}. \tag{3}
\end{equation*}

Standard results from estimation theory allow the MMSE channel estimate between of $\boldsymbol{h}_{k}^{[n]}$ to be expressed as

\begin{equation*}
\hat{\boldsymbol{h}}_{k}^{[n]}=\left(\sqrt{\tau_{p}\left(P_{p}^{\mathrm{AP}}\right)} / \sigma_{u}^{2}\right) \boldsymbol{R}_{n k} \boldsymbol{\Psi}_{n k}^{-1} \breve{\boldsymbol{y}}_{p k}^{[n]}, \tag{4}
\end{equation*}
where
\begin{equation*}
\boldsymbol{\Psi}_{n k}=\tau_{p}\left(P_{p}^{\mathrm{AP}} / \sigma_{u}^{2}\right) \sum_{k^{\prime}=1}^{K} \boldsymbol{R}_{n k^{\prime}}\left|\boldsymbol{\varphi}_{k^{\prime}}^{H} \boldsymbol{\varphi}_{k}\right|^{2}+\boldsymbol{I}_{N}. \tag{5}
\end{equation*}

The channel estimate $\hat{\boldsymbol{h}}_{k^{[n]}}^{[n]}$ and the MMSE channel estimation error $\tilde{\boldsymbol{h}}_{k}^{[n]}=\boldsymbol{h}_{k}^{[n]}-\hat{\boldsymbol{h}}_{k}^{[n]}$ are uncorrelated random vectors distributed as $\hat{\boldsymbol{h}}_{k}^{[n]} \sim \mathcal{C N}\left(\mathbf{0}, \boldsymbol{\Gamma}_{m k}\right)$, and $\tilde{\boldsymbol{h}}_{k}^{[n]} \sim \mathcal{C N}\left(\mathbf{0}, \boldsymbol{A}_{m k}\right)$, respectively, where

\begin{equation*}
\boldsymbol{\Gamma}_{n k}=\tau_{p}\left(P_{p}^{\mathrm{AP}} / \sigma_{u}^{2}\right) \boldsymbol{R}_{n k} \boldsymbol{\Psi}_{n k}^{-1} \boldsymbol{R}_{n k}^{H} \tag{6}
\end{equation*}
and

\begin{equation*}
\boldsymbol{A}_{n k}=\mathbb{E}\left\{\tilde{\boldsymbol{h}}_{k}^{[n]} \tilde{\boldsymbol{h}}_{k}^{[n], H}\right\}=\boldsymbol{R}_{n k}-\boldsymbol{\Gamma}_{n k}. \tag{7}
\end{equation*}

\subsection{SINR computation}
We aim to optimize the downlink transmission, assuming perfect channel state information. This is, the downlink transmission is performed by a central controller which has a perfect estimation of $\mathbf{h}_{k} \in \mathbb{C}^{N N_{\mathrm{AP}} \times 1}$ for $k=1, \ldots, K$. Vector $\mathbf{h}_{k}$ collapses the channel vector of each AP

\begin{equation*}
\mathbf{h}_{k}=\left(\mathbf{h}_{k}^{[1], T}, \ldots, \mathbf{h}_{k}^{[N], T}\right)^{T}, \tag{8}
\end{equation*}

where $\mathbf{h}_{k}^{[n], T} \in \mathbb{C}^{N_{\mathrm{AP}} \times 1}$ is the channel vector between the $n$-th AP and the $k$-th UE.

The received signal by the $k$-th UE can be written as

\begin{equation*}
y_{k}=\mathbf{h}_{k}^{H} \mathbf{w}_{k} s_{k}+\sum_{j \neq k}^{K} \mathbf{h}_{k}^{H} \mathbf{w}_{j} s_{j}+n_{k}, \tag{9}
\end{equation*}

where $s_{k}$ is the unit norm zero mean symbol of the $k$-th UE and $\mathbf{w}_{k} \in \mathbb{C}^{N N_{\mathrm{AP}} \times 1}$ the transmit precoding vector. Note that transmission takes place in all APs simultaneously so that the $n$-th AP is transmitting

\begin{equation*}
\mathbf{t}^{[n]}=\sum_{k=1}^{K} \mathbf{w}_{k}^{[n]} s_{k}, \tag{10}
\end{equation*}
where $\mathbf{w}_{k}^{[n]} \in \mathbb{C}^{N_{\mathrm{AP}} \times 1}$ is the precoding vector at the $n$-th AP transmitting the symbol to the $k$-th UE. Remarkably

\begin{equation*}
\mathbf{w}_{k}=\left(\mathbf{w}_{k}^{[1], T}, \ldots, \mathbf{w}_{k}^{[N], T}\right)^{T}. \tag{11}
\end{equation*}
We consider total power constraint

\begin{equation*}
\sum_{k=1}^{K}\left\|\mathbf{w}_{k}^{[n]}\right\|^{2} \leq P_{n}, \quad n=1, \ldots, N \tag{12}
\end{equation*}
where $P_{n}$ is the available power at the $n$-th AP. The attainable rate of the $k$-th user is

\begin{equation*}
R_{k}=\log _{2}\left(1+\operatorname{SINR}_{k}\right), \tag{13}
\end{equation*}
where

\begin{equation*}
\operatorname{SINR}_{k}=\frac{\left|\mathbf{h}_{k}^{H} \mathbf{w}_{k}\right|^{2}}{\sum_{j \neq i}^{K}\left|\mathbf{h}_{k}^{H} \mathbf{w}_{j}\right|^{2}+\sigma^{2}}. \tag{14}
\end{equation*}
\bigskip

\section{Problem Statement}
In this scenario, we consider minimizing the global power consumption guaranteeing a certain minimum SINR, $\gamma_{0}$ to all terminals. We model the Total Power Consumption (TPC) as follows

\begin{equation*}
\mathrm{TPC}=\frac{1}{\eta} \sum_{k=1}^{K}\left\|\mathbf{w}_{k}\right\|^{2}+\sum_{n=1}^{N} P_{\mathrm{FIX}, n} \tag{15}
\end{equation*}
where $\eta$ is the high power amplifier efficiency, $\eta \in(0,1]$ assumed to be the same for all APs and all antennas. The term $P_{\mathrm{FIX}, n}$ denotes the fixed consumed power due to backhauling, circuitry power consumption, and control signaling of the $n$-th AP in case it is active. This power consumption model aligns with other related approaches to the energy consumption of CF-mMIMO systems \cite{9353695}, \cite{8097026}.

Mathematically, this can be written as
\begin{align*}
& \underset{\left\{\mathbf{w}_{k}\right\}_{k=1}^{K}, \mathcal{U}}{\operatorname{minimize}}    \frac{1}{\eta} \sum_{k=1}^{K}\left\|\mathbf{w}_{k}\right\|^{2}+P_{\mathrm{FIX}}|\mathcal{U}| \\
& \text { subject to } \\
& \sum_{k=1}^{K}\left\|\mathbf{w}_{k}^{[n]}\right\|^{2} \leq P_{n} \quad n=1, \ldots, N  \tag{16}\\
& \operatorname{SINR}_{k} \geq \gamma_{0} \quad k=1, \ldots, K \\
& \mathcal{U} \subseteq \mathcal{V},
\end{align*}
where $\mathcal{V}$ denotes the set with all the APs and $|\cdot|$ denotes the cardinality of the set. We assume all APs have the same fixed power consumption $P_{\text {FIX }}$. The aforementioned cardinality problem can be re-written as follows

$$
\underset{\left\{\mathbf{w}_{k}\right\}_{k=1}^{K},\left\{b_{n}\right\}_{n=1}^{N}}{\operatorname{minimize}} \frac{1}{\eta} \sum_{k=1}^{K}\left\|\mathbf{w}_{k}\right\|^{2}+P_{\mathrm{FIX}} \sum_{n=1}^{N} b_{n}
$$
 subject to
\begin{align*}
& C 1: \sum_{k=1}^{K}\left\|\mathbf{w}_{k}^{[n]}\right\|^{2} \leq b_{n} P_{n} \quad n=1, \ldots, N  \tag{17}\\
& C 2: \operatorname{SINR}_{k} \geq \gamma_{0} \quad k=1, \ldots, K \\
& C 3: b_{n} \in\{0,1\} \quad n=1, \ldots, N.
\end{align*}

This latter optimization problem is non-convex due to constraints $C 2$ and $C 3$. Let us focus on how to deal with these non-convex constraints. As reported in [14], constraint $C 1$ can be rewritten as

\begin{equation*}
\frac{1}{\gamma_{0}}\left|\mathbf{h}_{k}^{H} \mathbf{w}_{k}\right|^{2} \geq \sum_{j \neq k}^{K}\left|\mathbf{h}_{k}^{H} \mathbf{w}_{j}\right|^{2}+\sigma^{2} \tag{18}.
\end{equation*}

This last expression shows that an arbitrary phase shift does not affect the constraint. Therefore, we can consider a real solution such that its imaginary part  $\Im\left(\mathbf{h}_{k}^{H} \mathbf{w}_{k}\right)=0$. In this context, we can transform the original constraint into

\begin{equation*}
\frac{1}{\sqrt{\gamma_{0}}} \Re\left(\mathbf{h}_{k}^{H} \mathbf{w}_{k}\right) \geq \sqrt{\sum_{j \neq k}^{K}\left|\mathbf{h}_{k}^{H} \mathbf{w}_{j}\right|^{2}+\sigma^{2}}, \tag{19}
\end{equation*}
which is a convex constraint. Alternatively, $C 3$ is a binary constraint whose quadratic definition, as reported in \cite{8292945}, can be written as
\begin{equation*}
\mathbf{b}^{T} \mathbf{E}_{i} \mathbf{b}-\mathbf{b}^{T} \mathbf{e}_{i}=0 .\tag{20}
\end{equation*}
where $\mathbf{E}_{i}$ is a zero matrix whose $i$-th diagonal element is equal to 1 and $\mathbf{e}_{i}$ is a zero vector whose $i$-th entry is equal to one. This equality constraint can be divided into two different constraints, namely

\begin{align*}
& \mathbf{b}^{T} \mathbf{E}_{i} \mathbf{b}-\mathbf{b}^{T} \mathbf{e}_{i} \leq 0  \tag{21}\\
& \mathbf{b}^{T} \mathbf{E}_{i} \mathbf{b}-\mathbf{b}^{T} \mathbf{e}_{i} \geq 0 \tag{22}.
\end{align*}

While the first constraint is convex, the latter is concave. For this last constraint, we resort to the Concave-Convex Procedure (CCP), which sequentially approximates the concave parts of the problem by its first order Taylor expansion. Starting from an initial point $\mathbf{z}$, CCP algorithm linearizes the non-convex part, by linearizing it by its affine approximation, leading into the following constraint

\begin{equation*}
\mathbf{b}^{T} \mathbf{e}_{i}+\mathbf{z}^{T} \mathbf{E}_{i} \mathbf{z}-2 \Re\left(\mathbf{z}^{T} \mathbf{E}_{i} \mathbf{b}\right) \leq 0 \tag{23}.
\end{equation*}

Bearing in mind the above discussion and the use of slack variables $\mathbf{s}$, the original problem at the $t$-th step becomes

$$ \underset{\left\{\mathbf{w}_{k}\right\}_{k=1}^{K},\left\{b_{n}\right\}_{n=1}^{N},\left\{\mathbf{s}_{n}\right\}_{n=1}^{N}}{\operatorname{minimize}} \frac{1}{\eta} \sum_{k=1}^{K}\left\|\mathbf{w}_{k}\right\|^{2}+P_{\mathrm{FIX}} \sum_{b=1}^{N} b_{n}+\lambda \sum_{n=1}^{N} s_{n} $$

subject to

\begin{align*}
    C 1:& \quad \sum_{k=1}^{K}\left\|\mathbf{w}_{k}^{[n]}\right\|^{2} \leq b_{n} P_{n}, n=1, \ldots, N, \\
    C 2.1:& \quad \frac{1}{\sqrt{\gamma_{0}}} \Re\left(\mathbf{h}_{k}^{H} \mathbf{w}_{k}\right) \geq \sqrt{\sum_{j \neq k}^{K}\left|\mathbf{h}_{k}^{H} \mathbf{w}_{j}\right|^{2}+\sigma^{2}},\\ & 
                k=1, \ldots, K, \\
    C 2.2:& \quad \Im\left(\mathbf{h}_{k}^{H} \mathbf{w}_{k}\right)=0, k=1, \ldots, K, \\
    C 3.1:& \quad \mathbf{b}^{T} \mathbf{E}_{i} \mathbf{b}-\mathbf{b}^{T} \mathbf{e}_{i} \leq 0,  n=1, \ldots, N, \\
    C 3.2:& \quad \mathbf{b}^{T} \mathbf{e}_{i}+\mathbf{z}_{t}^{T} \mathbf{E}_{i} \mathbf{z}_{t}-2 \Re\left(\mathbf{z}_{t}^{T} \mathbf{E}_{i} \mathbf{b}\right) \leq s_i. \tag{24}
\end{align*}

The penalized concave-convex procedure is summarized in Algorithm 1. Note that slack variables $s_i$ are introduced to avoid the need for an initial feasible solution. In our case, we consider as initial solution the one that uses all APs.

\begin{algorithm}
\caption{PCCP optimization for On/Off Optimization}
\begin{algorithmic}[1]
\algnotext{EndIf}
\algnotext{EndWhile}
\State \textbf{Initialization} of $z^{(0)}$,$b^{(0)},w^{(0)}$ to random initial points.
\bigskip
\State \textbf{Set} $t=0$ and define the values of $\psi, T_{max}, \omega,\lambda_{\text{max}}$, $\rho $ and $\lambda^{(0)}$
\bigskip
\While{$\sum_{m=1}^{2N+1} s_m \leq \psi$ and $\|b^{(n)} - b^{(n-1)}\| \geq \omega$}
    \If{$t < T_{\text{max}}$}
        \State Compute $b^{(n)}$ according to (24).
        \State $z^{(n+1)} \leftarrow b^{(n)}$;
        \State $\lambda^{(n+1)} \leftarrow \max (\lambda^{(n)} \rho, \lambda_{\text{max}})$;
        \State $t \leftarrow t + 1$;
    \Else
        \State $t \leftarrow 0$;
        \State Initialize with a new random value $z^{(0)}$;
        \State Set up $\lambda^{(0)}$ again;
    \EndIf
\EndWhile
\State {Output the final solution $\mathbf{w^*,b^*}$;}
\end{algorithmic}
\end{algorithm}

\section{Numerical Results}



This section quantifies the performance of the joint AP selection and power consumption minimization algorithm in a cell-free mMIMO O-RAN scenario. 

The power saving \ac{xApp} is placed in Near-RT-RIC, as shown in Figure 1. We consider $N$=15 APs with four antennas each. The channels are generated following a complex normal distribution of zero mean and unit variance. The number of users $K$ is considered to be 10. The maximum transmit power is fixed to 1W, maximum value of penalty parameter $\lambda_{max}$ is fixed to $10^4$, maximum allowed constraints violation $\psi$ is $10^{-5}$, initial value $\lambda^{(0)}$ is set as $5\cdot 10^{-2}$, multiplicative update value $\rho$ is fixed to 2, maximum number of iterations $T_{max}$ is $10^2$ and finally $10^{-5}$ is assigned to maximum allowed constraints violation $\omega$. All the experiments are run for 100 Monte Carlo runs.

\begin{figure}[htp!]
\centering
\includegraphics[width=1\linewidth, height=0.23\textheight]{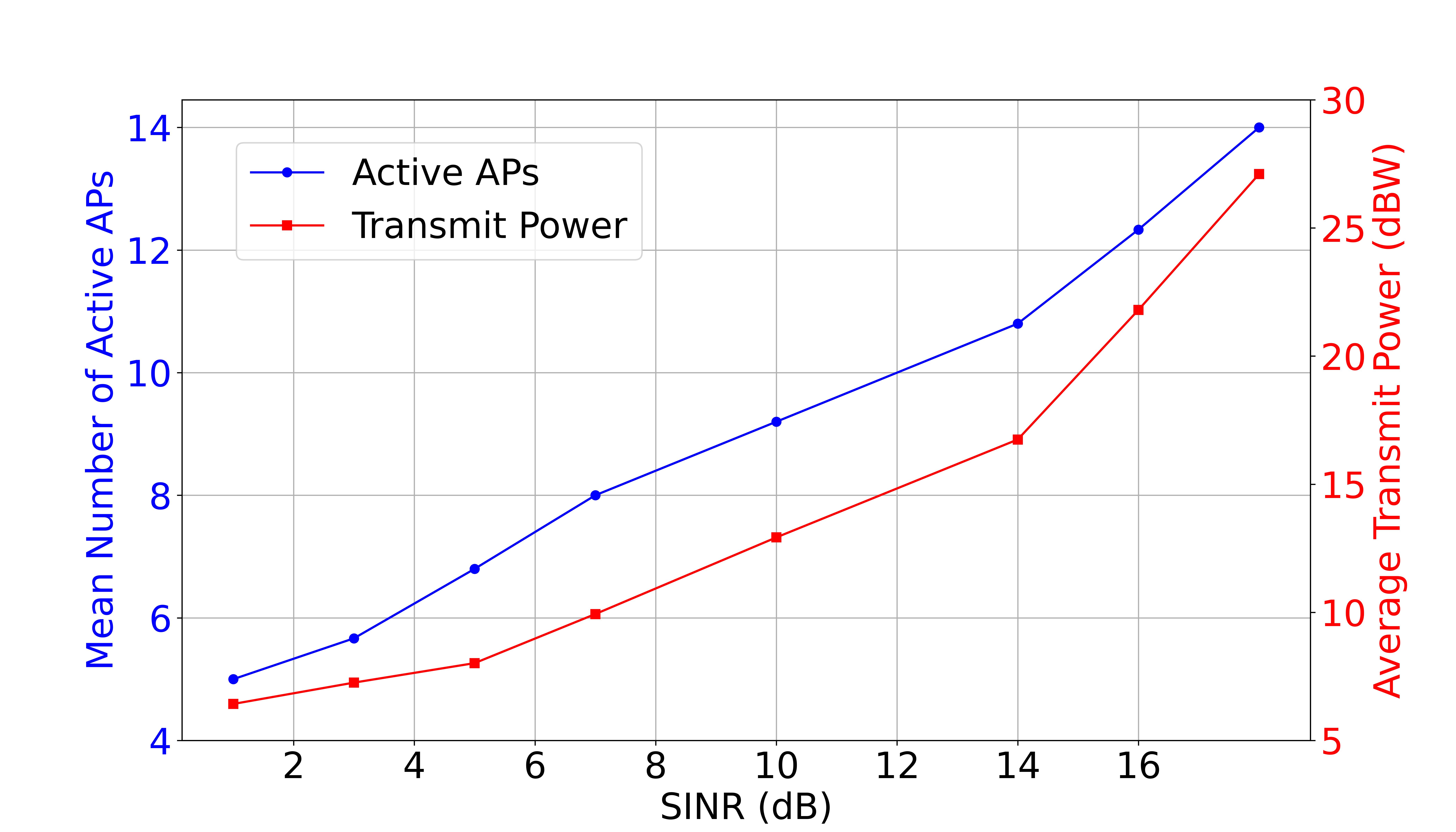}
\caption{Mean number active APs and average transmit power as a function of SINR.}
\label{APs}
\end{figure}

Fig. \ref{APs} helps understand the relationship between SINR, number of active APs, and power consumption. Our work aims to jointly optimize total power consumption and beamforming design. A direct consequence of solving the problem (24) is to select the best set of access points to be activated. The results show that the mean number of active access points monotonically increases with SINR. An increase in SINR signifies better signal quality at the UEs, requesting demands for more active access points, leading to a gradual increase in power consumption. The results show that up to 7dB to fulfill the SINR constraints implies a slight increase in the total power. On the contrary, for values higher than 15dB, there is a huge increase in total transmit power, around 9dB, to guarantee the quality of service. This analysis highlights the significant tradeoff between network performance in terms of UE's quality of service and energy efficiency, which, as a consequence, results in increasing active APs. 
\begin{figure}[htp!]
\centering
\includegraphics[width=\linewidth, height=0.23\textheight]{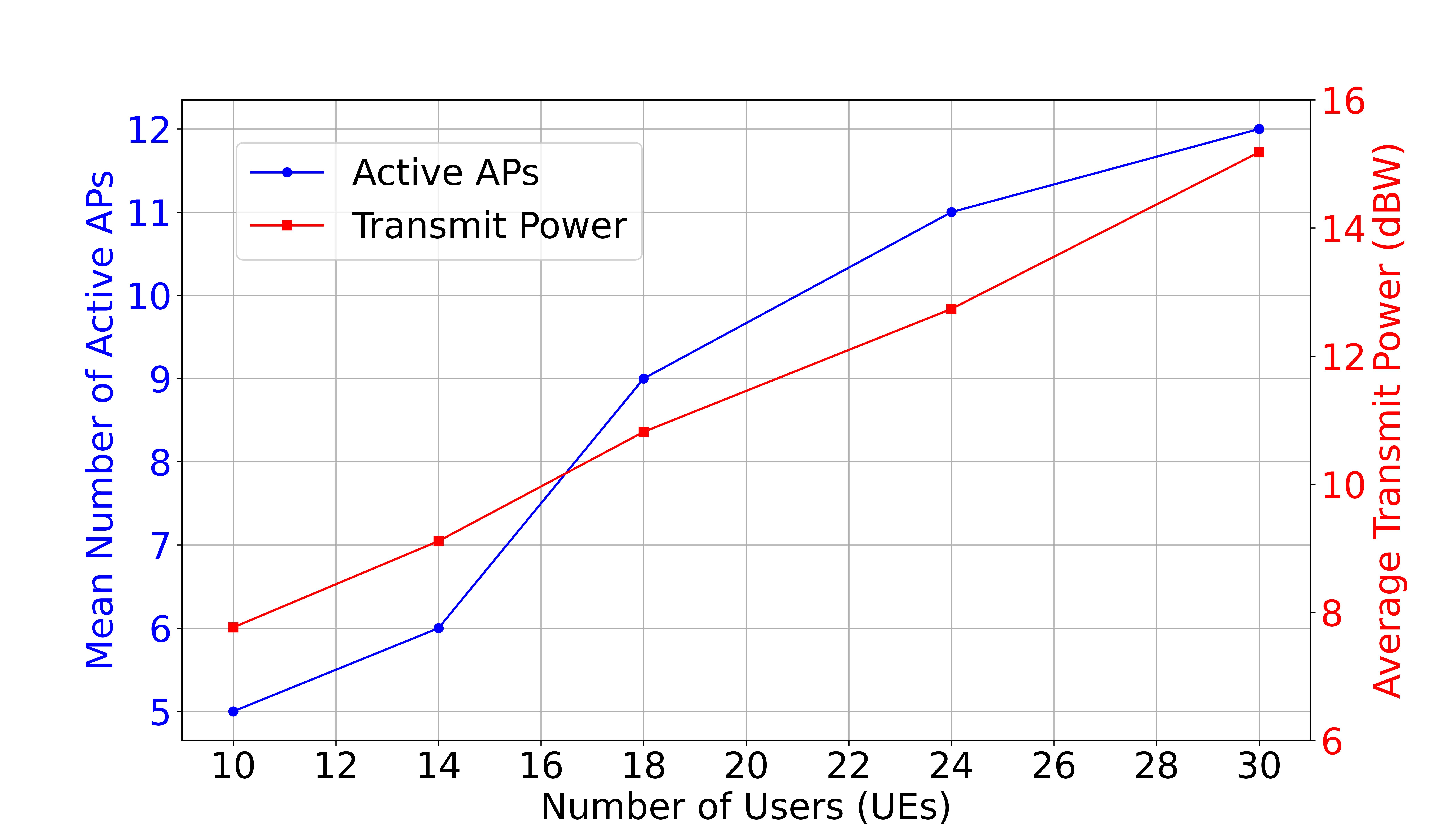}
\caption{Effect of changing the number of users on active APs and transmit power.}
\label{UEs}
\end{figure}

Fig. \ref{UEs} shows the impact of increasing the number of UEs on the average transmit power and the mean number of active access points. For this analysis, the SINR value was fixed to 4dB. As the total number of UEs rises, there is a consistent increase in the number of APs activated to serve these users. In the same way, the transmit power is also increased. To maintain the quality of service, in terms of SINR at the terminals, a higher number of active points needs to be increased, having an impact on the total transmit power in the system. 

\begin{figure}[htp!]
\centering
\includegraphics[width=\linewidth]{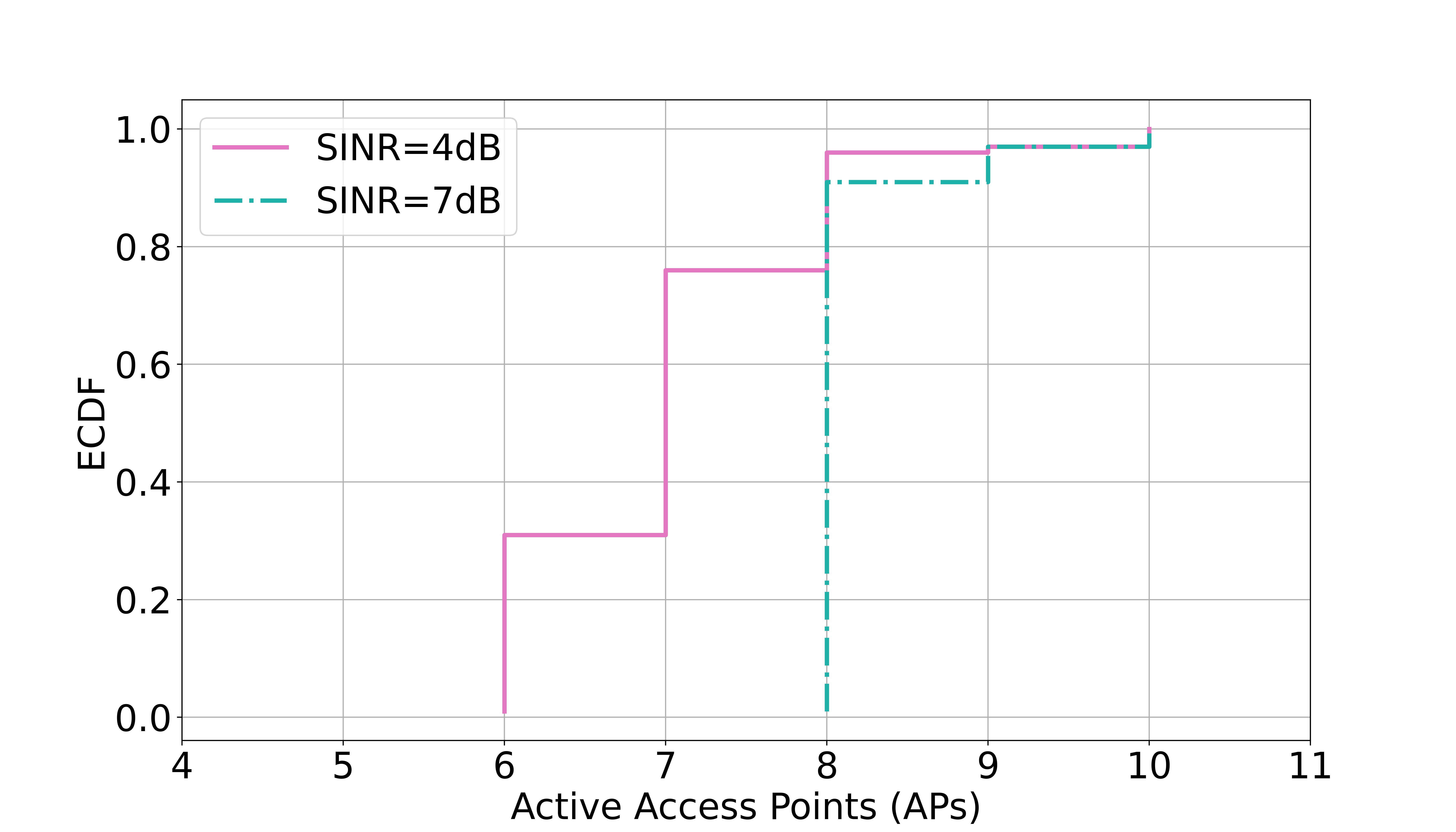}
\caption{Empirical Cumulative Distribution Function of the number of active APs for different SINRs.}
\label{ECDF}
\end{figure}

The Fig. \ref{ECDF} shows the Empirical Cumulative Distribution Function (ECDF) of the actual number of active APs corresponding to 100 Montecarlo runs for SINRs equal to 4dB and 7dB, the number of available APs $N=10$, and the number of UEs $K=21$. More access points need to be activated as the SINR constraint increases, shifting the ECDFs of required APs further to the right. This is because to find a feasible solution in higher SINR regimes, a larger minimum amount of APs is needed, as shown in the 7dB case. Around 90\% of cases require 8 APs to fulfill the constraints. Furthermore, the figure shows that for lower values of SINR, i.e., 4dB, broader solutions in terms of active APs are found. For this case, the total amount of APs required ranges from 6 to 10.

\begin{figure}[htp!]
\centering
\includegraphics[width=\linewidth, height=0.23\textheight]{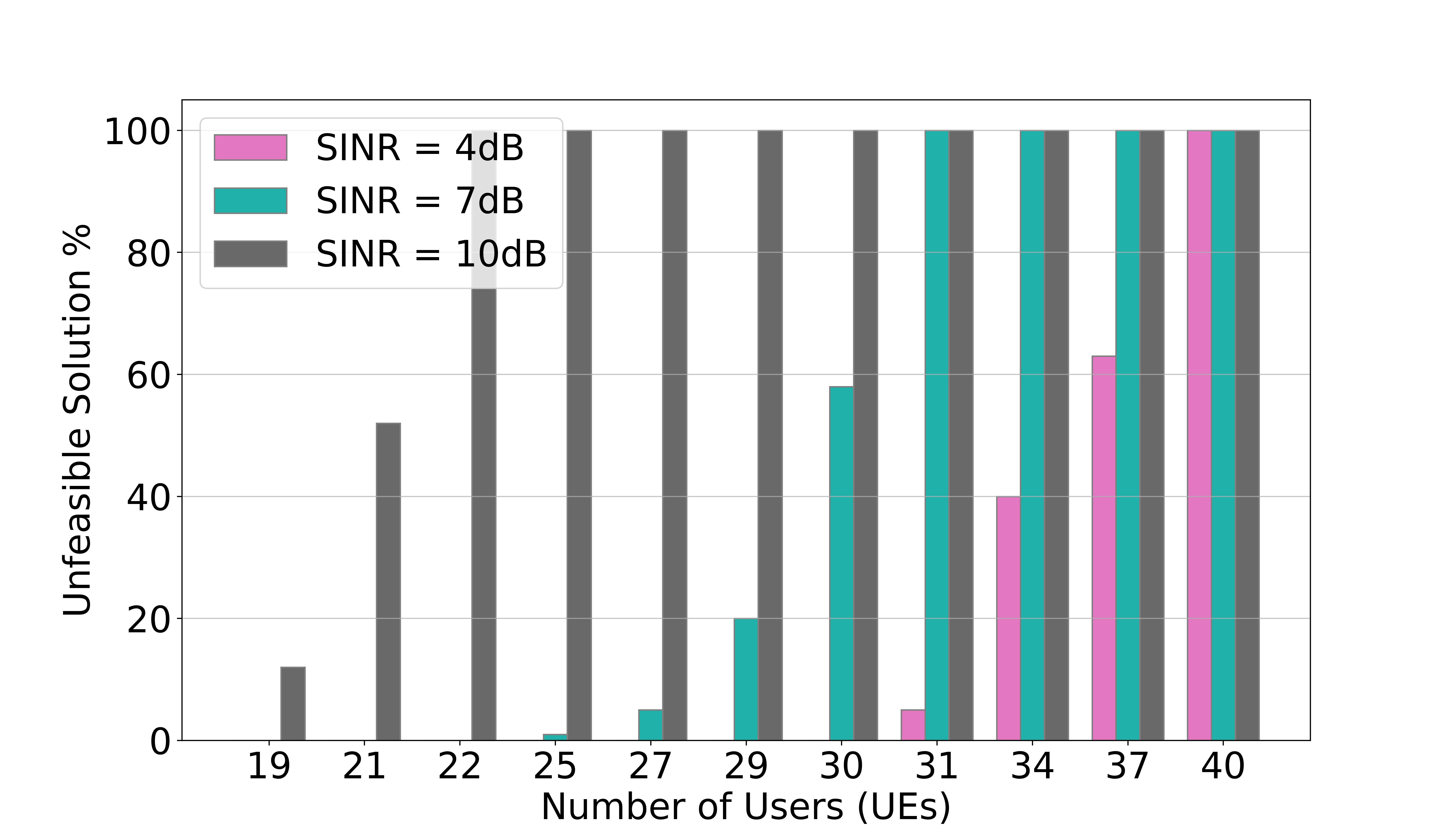}
\caption{Feasibility of finding an optimal solution for various SINRs.}
\label{Ufs}
\end{figure}

Fig. \ref{Ufs} investigates the effect of increasing the total number
of users on the feasibility of finding optimal solutions to
problem (24). Having this aim in mind, different SINR values were evaluated, namely 4dB, 7dB, and 10dB, with the number of available APs $N=10$. In the case of 4dB SINR, when the number of users in the system is higher than 30, in some instances the problem becomes unfeasible. Similarly, for cases of 7dB and 10dB, this unfeasibility effect starts earlier when the number of users is 25 and 19, respectively. The rationale behind this behavior is that for the given number of APs and antennas, $N=10$ and $K=4$, respectively, fulfilling the quality of service constraints becomes unfeasible for a higher number of users. This effect is consistently increased, and when the number of users reaches 40, the optimization problem completely fails in finding a feasible solution for all the listed SINRs. Note that increasing the number of APs and/or antennas would allow for a higher number of users to be served before reaching unfeasibility. 



\section{Conclusions}
This paper presented a novel approach for addressing the energy consumption minimization problem in Open RAN CF-mMIMO systems while guaranteeing the quality of service requirements per user. This problem involves the joint precoding design and the AP selection, with a minimum SINR requirement for each user. The combinatorial nature of the problem makes it computationally demanding, and a sub-optimal method was presented, based on the penalized convex-concave procedure. Our technique offered a promising solution, demonstrating its efficacy through numerical simulations. Furthermore, the proposed approach, which can be implemented as an xApp in the near-time RIC, holds significant potential for practical implementation of O-RAN-based CF-mMIMO systems.

\bibliographystyle{ieeetr}
\bibliography{biblio}  

\begin{thebibliography}{10}

\bibitem{9356519}
M.~Matthaiou, O.~Yurduseven, H.~Q. Ngo, D.~Morales-Jimenez, S.~L. Cotton, and V.~F. Fusco, ``The road to 6g: Ten physical layer challenges for communications engineers,'' {\em IEEE Communications Magazine}, vol.~59, no.~1, pp.~64--69, 2021.

\bibitem{7827017}
H.~Q. Ngo, A.~Ashikhmin, H.~Yang, E.~G. Larsson, and T.~L. Marzetta, ``Cell-free massive mimo versus small cells,'' {\em IEEE Transactions on Wireless Communications}, vol.~16, no.~3, pp.~1834--1850, 2017.

\bibitem{10437450}
R.~Beerten, V.~Ranjbar, A.~P. Guevara, and S.~Pollin, ``Cell-free massive mimo in the o-ran architecture: Cluster and handover strategies,'' in {\em IEE GLOBECOM}, pp.~5943--5948, 2023.

\bibitem{9353695}
A.~Papazafeiropoulos, H.~Q. Ngo, P.~Kourtessis, S.~Chatzinotas, and J.~M. Senior, ``Towards optimal energy efficiency in cell-free massive mimo systems,'' {\em IEEE Transactions on Green Communications and Networking}, vol.~5, no.~2, pp.~816--831, 2021.

\bibitem{femenias2020access}
G.~Femenias, N.~Lassoued, and F.~Riera-Palou, ``Access point switch on/off strategies for green cell-free massive mimo networking,'' {\em IEEE access}, vol.~8, pp.~21788--21803, 2020.

\bibitem{Demir_CF_ORAN}
{\"O}.~T. Demir, M.~Masoudi, E.~Bj{\"o}rnson, and C.~Cavdar, ``Cell-free massive mimo in o-ran: Energy-aware joint orchestration of cloud, fronthaul, and radio resources,'' {\em IEEE Journal on Selected Areas in Communications}, 2024.

\bibitem{zhou2020max}
A.~Zhou, J.~Wu, E.~G. Larsson, and P.~Fan, ``Max-min optimal beamforming for cell-free massive mimo,'' {\em IEEE Communications Letters}, vol.~24, no.~10, pp.~2344--2348, 2020.

\bibitem{lipp2016}
T.~Lipp and S.~Boyd, ``Variations and extension of the convex--concave procedure,'' {\em Optimization and Engineering}, vol.~17, pp.~263--287, 2016.

\bibitem{10266607}
V.~Kasuluru, L.~Blanco, and E.~Zeydan, ``On the use of probabilistic forecasting for network analysis in open ran,'' in {\em 2023 IEEE Meditcom}, pp.~258--263, 2023.

\bibitem{polese2023understanding}
M.~Polese {\em et~al.}, ``Understanding o-ran: Architecture, interfaces, algorithms, security, and research challenges,'' {\em IEEE Communications Surveys \& Tutorials}, 2023.

\bibitem{kliks2023towards}
A.~Kliks, M.~Dryjanski, V.~Ram, L.~Wong, and P.~Harvey, ``Towards autonomous open radio access networks,'' {\em ITU Journal on Future and Evolving Technologies}, vol.~4, no.~2, pp.~251--268, 2023.

\bibitem{bonati2021intelligence}
L.~Bonati, S.~D'Oro, M.~Polese, S.~Basagni, and T.~Melodia, ``Intelligence and learning in o-ran for data-driven nextg cellular networks,'' {\em IEEE Communications Magazine}, vol.~59, no.~10, pp.~21--27, 2021.

\bibitem{10330597}
M.~S. Oh, A.~B. Das, S.~Hosseinalipour, T.~Kim, D.~J. Love, and C.~G. Brinton, ``A decentralized pilot assignment algorithm for scalable o-ran cell-free massive mimo,'' {\em IEEE Journal on Selected Areas in Communications}, 2023.

\bibitem{li2017intelligent}
R.~Li, Z.~Zhao, X.~Zhou, G.~Ding, Y.~Chen, Z.~Wang, and H.~Zhang, ``Intelligent 5g: When cellular networks meet artificial intelligence,'' {\em IEEE Wireless communications}, vol.~24, no.~5, pp.~175--183, 2017.

\bibitem{8097026}
H.~Q. Ngo, L.-N. Tran, T.~Q. Duong, M.~Matthaiou, and E.~G. Larsson, ``On the total energy efficiency of cell-free massive mimo,'' {\em IEEE Transactions on Green Communications and Networking}, vol.~2, no.~1, pp.~25--39, 2017.

\bibitem{8292945}
M.~A. V{\'a}zquez, L.~Blanco, and A.~I. P{\'e}rez-Neira, ``Hybrid analog--digital transmit beamforming for spectrum sharing backhaul networks,'' {\em IEEE transactions on signal processing}, vol.~66, no.~9, pp.~2273--2285, 2018.

\end{thebibliography}

\end{document}